\documentclass[12pt,aps,prd,preprint,tightenlines,superscriptaddress,nofootinbib]{revtex4-1}
\pdfoutput=1
\usepackage{graphicx}
\usepackage{amsmath,amssymb,mathrsfs}
\usepackage{bbm}
\usepackage{color}
\usepackage{dsfont}
\usepackage{cancel}

\newcommand{\PRE}[1]{{#1}} 

\newcommand{\ben}{\begin{enumerate}}
\newcommand{\een}{\end{enumerate}}
\newcommand{\bit}{\begin{itemize}}
\newcommand{\eit}{\end{itemize}}

\newcommand{\beqa}{\begin{eqnarray}}
\newcommand{\eeqa}{\end{eqnarray}}
\newcommand{\beq}{\begin{equation}}
\newcommand{\eeq}{\end{equation}}
\newcommand{\bay}{\begin{array}}
\newcommand{\eay}{\end{array}}

\def\ifmath#1{\relax\ifmmode #1\else $#1$\fi}

\def\gsim{\ \rlap{\raise 3pt \hbox{$>$}}{\lower 3pt \hbox{$\sim$}}\ }
\def\lsim{\ \rlap{\raise 3pt \hbox{$<$}}{\lower 3pt \hbox{$\sim$}}\ }

\def\ls#1{\ifmath{_{\lower1.5pt\hbox{$\scriptstyle #1$}}}}
\def\lsup#1{^{\lower 6pt\hbox{$\scriptstyle#1$}}}

\def\ord{{\cal O}}
\def\lt{\left}
\def\rt{\right}

\def\rank#1{{{\rm rank} #1}}

\def\rkR{{\rank(R)}}

\def\detR{{\det R}}

\def\bracket#1#2 {\mathinner{\langle{#1}|{#2}\rangle}}

\def\eg{{\it e.g.}}
\def\ie{{\it i.e.}}

\def\bracket#1#2 {\mathinner{\langle{#1}|{#2}\rangle}}

\newcommand{\Eqref}[1]{Eq.~(\ref{eq:#1})}
\newcommand{\secref}[1]{Sec.~\ref{sec:#1}}

\newcommand{\tabref}[1]{Table~\ref{tab:#1}}

\newcommand{\ifb}{\text{fb}^{-1}}

\renewcommand{\vec}[1]{\mathbf{#1}}
\renewcommand{\vec}[1]{\mbox{\boldmath${#1}$}}
 
\begin{document}

\hfill YITP-SB-13-01

\vspace*{1cm}

\title{How to test for mass degenerate Higgs resonances\PRE{\vspace*{.2in}}}

\author{Yuval Grossman\PRE{\vspace*{.2in}}}
\affiliation{Laboratory for Elementary-Particle Physics, Cornell University, Ithaca, N.Y.
\PRE{\vspace*{.2in}}
}

\author{Ze'ev Surujon}
\affiliation{C. N. Yang Institute for Theoretical Physics, 
Stony Brook University, Stony Brook, NY 11794
\PRE{\vspace*{0.2in}}
}

\author{Jure Zupan}
\affiliation{Department of Physics, University of Cincinnati, Cincinnati, OH 45221
\PRE{\vspace*{0.4in}}
}

\begin{abstract}
\PRE{\vspace*{.3in}}
The Higgs-like signal observed at the LHC could be due to several mass
degenerate resonances. We show that the number of resonances is
related to the rank of a ``production and decay'' matrix,
$R_{if}$. Each entry in this matrix contains the observed rate in a
particular production mode $i$ and final state $f$.  In the case of
$N$ non-interfering resonances, the rank of $R$ is, at most, $N$. If
interference plays a role, the maximum rank is  generically $N^2$, or with a universal
phase, $N(N+1)/2$.  As an
illustration we use the present experimental data to constrain the rank
of the corresponding matrix.  We estimate the LHC reach of probing
two and three resonances under various
speculations on future measurements and uncertainties.
\end{abstract}

\maketitle

\section{Introduction}

At present there is clear evidence for a new boson at the Large
Hadron Collider (LHC) with a mass of about 126 GeV~\cite{:2012gk,:2012gu}.
In this paper we are interested in the possibility that the signal is
due to two or more mass degenerate states. 
We are especially interested in the question whether one can tell experimentally if there are more resonances, and perhaps even constrain their number, 
without actually resolving the resonant peaks.
That is, we are interested in the case that the resonances are mass degenerate to better than the experimental resolution.

So far, the observed rates $R_{if}$ of the 126 GeV resonance have been
consistent with the Standard Model (SM) Higgs boson, although their central values
do exhibit $\ord(1)$ deviations.  Here $i$ runs over the production modes: gluon fusion, vector boson fusion, associated production,  while the final states
$f$ are $\gamma\gamma$, $\tau\tau$, etc.
By itself, any deviation from the SM prediction in one element of $R_{if}$ can be
explained by some kind of a new physics scenario with one resonance.
Some patterns of deviations in $R_{if}$, however, are bound to be inconsistent
with the hypothesis of a single resonance, even if one allows for non-SM couplings.
This is easily understood using a counting argument.
We denote the number of observed production (decay) modes by $n_i$ $(n_f)$.
Assuming there is a single resonance, the unknowns are the $n_i$ cross sections
$\{\sigma_i\}$ and $n_f$ branching ratios $\{B_f\}$. The number of
observables is $n_in_f$, which are all the entires in $R_{if}$. Thus,
the set of equations one has to solve is
\beq
  \sigma_i B_f=R_{if}.
   \label{eq:single}
\eeq
In this system of equations the number of unknowns,  $n_i+n_f$,
can be less than the number of observables, $n_in_f$. 
With enough measurements, this may lead to a system of equations which is
overdetermined and potentially inconsistent.\footnote{%
For a single resonance, by taking the logarithm of each equation in \eqref{eq:single} one arrives at a system of linear equations 
with $\log\sigma_i$ and $\log B_f$ as the new variables.
Therefore, discussing linear dependence, over- and under-determination,
and (in)consistency is formally justified.
}
Such inconsistency cannot be removed by any type of new physics that affects only the production or
decay of a single resonance. We would then
be forced to abandon the single resonance hypothesis and conclude that there are
more resonances.

The idea of using such inconsistency for detecting two resonances has been pioneered in Ref.~\cite{Gunion:2012he},
where the authors used ratios of the form $R_{ia}/R_{ib}$. For fixed $a$ and $b$ these ratios are independent
of $i$, if there is only one resonance. This can then be used to test for the existence of more than one resonance
(invoking two resonances to enhance the diphoton rate has been mentioned recently in~\cite{Gunion:2012gc}
for the NMSSM and in~\cite{Ferreira:2012nv,Drozd:2012vf} for two Higgs doublet models).
In this paper we generalize the condition above to including any number of resonances, as well as interference effects.   Alternatively, the existence of extra resonances may be directly probed using line shapes, see e.g. 
\cite{Ellis:2005fp,Ellis:2004fs}.

Assuming no interference between the resonances we find that $\rkR$, 
the rank of the signal matrix $R_{if}$, constitutes a lower bound
on the number of resonances required to explain the data.
For interfering resonances we find, on the other hand, that this lower bound is relaxed
to the integer that is greater or equal to $\sqrt{\rkR}$ in the most
general case or to $\big(\sqrt{1+8\,\rkR}-1\big)/2$ if a universal phase
(i.e. independent of $i$ and $f$) is assumed.
It follows that if $\rkR\geq2$, there must be at least two resonances,
whereas if $\rkR\geq3$, the resonances must interfere, or there
must be at least three of them.

\section{General discussion}
\label{sec:infinite}

We begin by recalling a few useful mathematical relations.
Consider two sets of vectors in $\mathbb{C}^N$ 
\beq
   \vec{u}^r\equiv\lt(u_1^r,\ldots,u_{N}^r\rt),\qquad
   \vec{v}^r\equiv\lt(v_1^r,\ldots,v_{N}^r\rt),
\eeq
where $r=1,\dots,n$ and we assume $N\gg n$.
We define two $N\times N$ matrices whose elements are
\beq
   A_{ij} = \sum_{r=1}^n \left|u_i^r \,v_j^r\right|^2, \qquad
   B_{ij} = \left|\sum_{r=1}^n u_i^r \, v_j^r\right|^2.
\eeq
Then one can prove that generically,
\beq \label{eq:mathrank}
   \rank\,A = n, \qquad \rank\,B =n^2.
\eeq
When the phase between $u_i^r v_j^r$ and $u_i^{r'} v_j^{r'}$ is independent of $i,j$,
the above results are modified
such that 
\beq \label{eq:mathrankreal}
   \rank\,A = n, \qquad \rank\,B =\frac{n(n+1)}{2}.
\eeq
Note that the condition leading to \eqref{eq:mathrankreal} is satisfied when the vectors are
real.

More generally, $u$ and $v$ can have different dimensions, and we may
have cancellations if the elements are not generic. Thus a more general
result is to treat the r.h.s. of the relations in 
Eqs.~(\ref{eq:mathrank}) and (\ref{eq:mathrankreal})
and as an
upper bound on the corresponding ranks. Of course, the rank is also
bounded by $N$.
Equipped with these results we now return to physics.

Suppose there exist $n_r>1$ resonances $\{\phi_1,\cdots,\phi_{n_r}\}$ all with masses close enough so that the experimental resolution is
not sufficient to resolve them.
If the masses of different resonances are split by much more than
their respective decay widths, interference effects are highly
suppressed and can be neglected. The signal strengths $R_{if}$ are
then given by 
\beq 
R_{if}=\sum_{r=1}^{n_r}\hat \sigma_i^r \hat B_f^r\equiv 
\vec{\hat \sigma}_i\cdot\vec{\hat B}_f,
   \label{eq:R_nonint}
\eeq
where the production cross sections $\hat \sigma_i^r\equiv
\sigma_i^r/\sigma_i^{\rm SM}$ and branching ratios $\hat B_i^r\equiv
B_i^r/B_i^{\rm SM}$ for each resonance $r$ are normalized to the SM 
values. The sum in \Eqref{R_nonint} runs over the degenerate
resonances $r=1,\dots,n_r$.
Using \Eqref{mathrank} we see that
\beq \label{eq:rank}
   n_r\geq\rkR.
\eeq
Namely, if one is able to measure $\rkR$, it would imply a lower
bound on the number of resonances.
Recall that the rank of a matrix $R_{if}$ is equal to the number of linearly
independent rows or columns, and may be obtained, \eg, using
singular value decomposition (SVD), where the number of nonzero singular
values equals $\rkR$. 

If the decay widths are larger than the mass splittings between
resonances, interference effects may be important.
For instance, for perfectly overlapping resonances,
\beq \label{eq:R_int}
   R_{if}=\lt|\sum_{r=1}^{n_r}A_{i}^r A_{f}^r\rt|^2
   \equiv\lt|\vec{A}_{i}\cdot\vec{A}_{f}\rt|^2,
\eeq
where $A_{i}^r$ and $A_{f}^r$ are the properly normalized amplitudes for
the production and decay of the resonance $r$, such that $\hat
\sigma_i^r=|A_i^r|^2$ and $\hat B_f^r=|A_f^r|^2$.  In contrast to the
non interfering case, the elements of $R_{if}$ are no longer
given by a sum of positive terms, and cancellations may occur.  Thus,
small entries in $R_{if}$ are possible, even when $\sigma_i^r$ and
$B_f^r$ are large, providing more model building freedom.

Moreover, interference affects the relation between $n_r$ and the rank of $R$.
Indeed, applying \Eqref{mathrank} to \Eqref{R_int} results in
\beq
   \rkR\leq n_r^2 \quad \Rightarrow \quad n_r \geq \sqrt{\rkR}.
   \label{eq:R_int2-b}
\eeq
In the case of a universal phase, i.e., a phase which depends only on the
resonance $\phi_r$ but not on the individual production ($i$)
and decay ($f$) mechanisms, we can apply \Eqref{mathrankreal} to
\Eqref{R_int} 
and obtain
\beq
   \rkR\leq \frac{n_r(n_r+1)}{2}.
   \label{eq:R_int2-b-cp}
\eeq
Solving for $n_r$ we obtain the lower bound
\beq
   n_r \geq \frac{1}{2}\big(\sqrt{1+8\,\rkR}-1\big).
\eeq
This occurs, for example, in the case where CP is conserved and the contributions of light fields running
in the gluon fusion and diphoton loops can be neglected. 
%
CP-even (``strong'') phases may come from three sources.
The first is the inherent Breit-Wigner $s$-dependence of the decay amplitude, which is
universal since it depends only on the decay width.
The second is rescattering of final states, which is not universal but can be neglected if
we assume perturbativity.
A third possible source of CP-even phase is when the leading diagram is a loop process
mediated by light fields running in the loop.
Then the imaginary component of the leading diagram may be sizable by virtue of the optical
theorem, resulting in an $\ord(1)$ non-universal phase.
This may occur, e.g., in certain large $\tan\beta$
scenarios, where gluon fusion or diphoton decay are mediated by the $b$ quarks and $\tau$
leptons.
%

It follows that any data which can be explained using $n_r$
non-interfering resonances may possibly also be explained with fewer
interfering ones.
For example, if the data satisfies $\rkR=3$, there must be
at least 2 interfering resonances, or 3 resonances without interference among
them. 
In the case where the data suggest $\rkR=4$, there must be
at least 2 interfering resonances or 3 interfering
resonances with universal phases or 4 non interfering ones.

We conclude this section with a few remarks.
\begin{enumerate}
\item
First we comment on the
transition between the interfering and non-interfering case. This transition must
be smooth, yet our result above seems discontinuous.
This may be understood as follows.
When interference is taken into account, the
extra non vanishing singular values of the matrix $R$ are proportional to the
size of the interference term.
When the resonances are far apart, their
interference is negligible and proportional to their overlap, and
the singular values approach zero. For example, consider two
resonances with equal width $\Gamma$ and mass difference $\Delta m$.
In this case, we expect two singular values to be similar, while the third one
is suppressed by the overlap.
In the Breit-Wigner approximation, it is given
by $(2\Gamma/\Delta m)^2$, although realistic profiles may deviate from the
Breit-Wigner curve at their tails.
\item
In case of universal phases, the interference can still increase the rank of $R$, however,
the resulting matrix is not generic.
For example, when all the amplitudes are real, the matrix $\sqrt{R_{ij}}$
has the same rank as that of $R_{ij}$ in the non interfering case.
\item
In all of the above, we have ignored effects of interference between the
signal and the SM background. Such effects were studied, for example, in
the case of $gg \to H \to \gamma\gamma$~\cite{Dixon:2003yb,Martin:2012xc}.
While this effect can be experimentally taken into account in the SM, it
cannot be done model independently.
Yet, taking the SM as a benchmark, we may estimate the effect to be
small enough, and ignore it. It may become an issue in the future.
\end{enumerate}

\section{Numerical Examples}
\label{sec:models}
We discuss next  a few concrete numerical examples, starting with the
information that can be extracted already with present experimental
precision and then contemplate projections to the future.  
The current status of the signal strength matrix $R_{if}$ is given in Table~\ref{tab:rates}, where we have averaged the experimental results from ATLAS and CMS.
%
\begin{table}
\centering
\begin{tabular}{cccccc}
   \hline\hline
prod.$\backslash$decay   & $\gamma\gamma$ & $WW^*$ & $\tau\bar\tau$ & $ZZ^*$  & $b\bar b$\\
   \hline
   $gg$ & $1.6\pm0.35$ \cite{ATLASgamma,CMScomb} & $0.8\pm0.3$ \cite{ATLAScomb,CMScomb} & $1.2\pm0.8$  \cite{ATLAStautau,CMStautau} & $1.0\pm0.3$ \cite{ATLASZZ}   & $-$\\
   VBF & $2.1\pm0.9$ \cite{ATLASgamma,CMScomb} & $-0.2\pm0.6$  \cite{ATLAScomb,CMScomb} & $0.3\pm0.7$ \cite{ATLAStautau,CMStautau} & $-$   & $-$\\
   $VH$ & $1.9\pm2.6$   \cite{ATLASgamma}& $-0.3\pm2.1$  \cite{CMScomb} & $1.0\pm1.8$ \cite{CMStautau} & $-$ & $0.8\pm0.6$ \cite{ATLASbbbar}\\
   $t\bar th$ & $-$ & $-$ & $-$ & $-$ & $<3.8$ \cite{CMSttbar} \\
   \hline\hline
\end{tabular}
\caption{The experimental values of signal strengths,
$R_{if}$,  for  several production modes (listed in the first column) and different final states (listed in the first row)  obtained by naively averaging ATLAS and CMS results with statistical and systematic errors added in quadrature and neglecting correlations. 
\label{tab:rates}}
\end{table}
%

In general, the signal strengths $R_{if}$
are measured with
finite precision $\delta R_{if}$. We simplify the discussion by neglecting
correlations between different channels, cross contaminations
and other systematic errors. 
For instance, depending on the cuts, some of the $gg$ signal can appear
in the vector boson fusion (VBF) production channel, as is the case in $h\to \gamma\gamma$. These effects were partially
taken into account in Table  \ref{tab:rates}, e.g. for the $\gamma\gamma$
final state. 
We also assume that the acceptances in the 
NP
models are similar to those of a SM Higgs boson.  This may be simply realized
by assuming that
different resonances are described by the same low energy effective
Lagrangian as the SM Higgs \cite{Espinosa:2012ir,Azatov:2012bz,Azatov:2012qz,Carmi:2012yp}.
While these approximations suffice at present precision,
we emphasize that these effects can be
important and there is no reason for them not to be included in a full-fledged analysis.

To constrain the number of resonances, we are interested in the
statistical properties of the hypothesis that $\rkR\geq r$. Note
that one can only put a lower bound on the rank since including experimental
errors generally increase it.
Finding the significances for each possible $r$ may be done in a
number of ways. Below we discuss two of these, but there are many
other possibilities. The best approach then depends on the particular situation.
Depending on the exact situation, one way may be
better than another.

\subsection{Using Singular Value Decomposition}
\begin{figure}
\begin{center}
\includegraphics[scale=0.6]{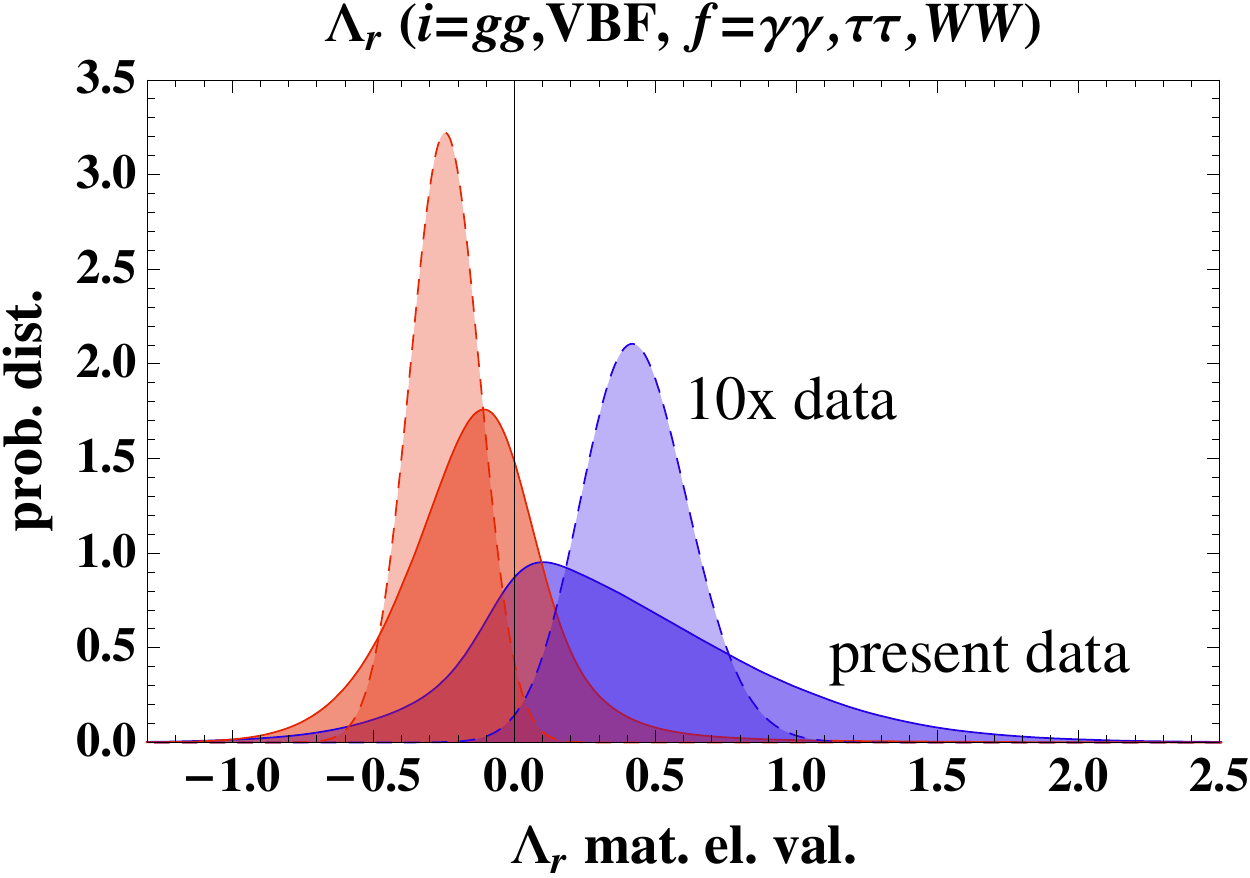}~~~~
\includegraphics[scale=0.6]{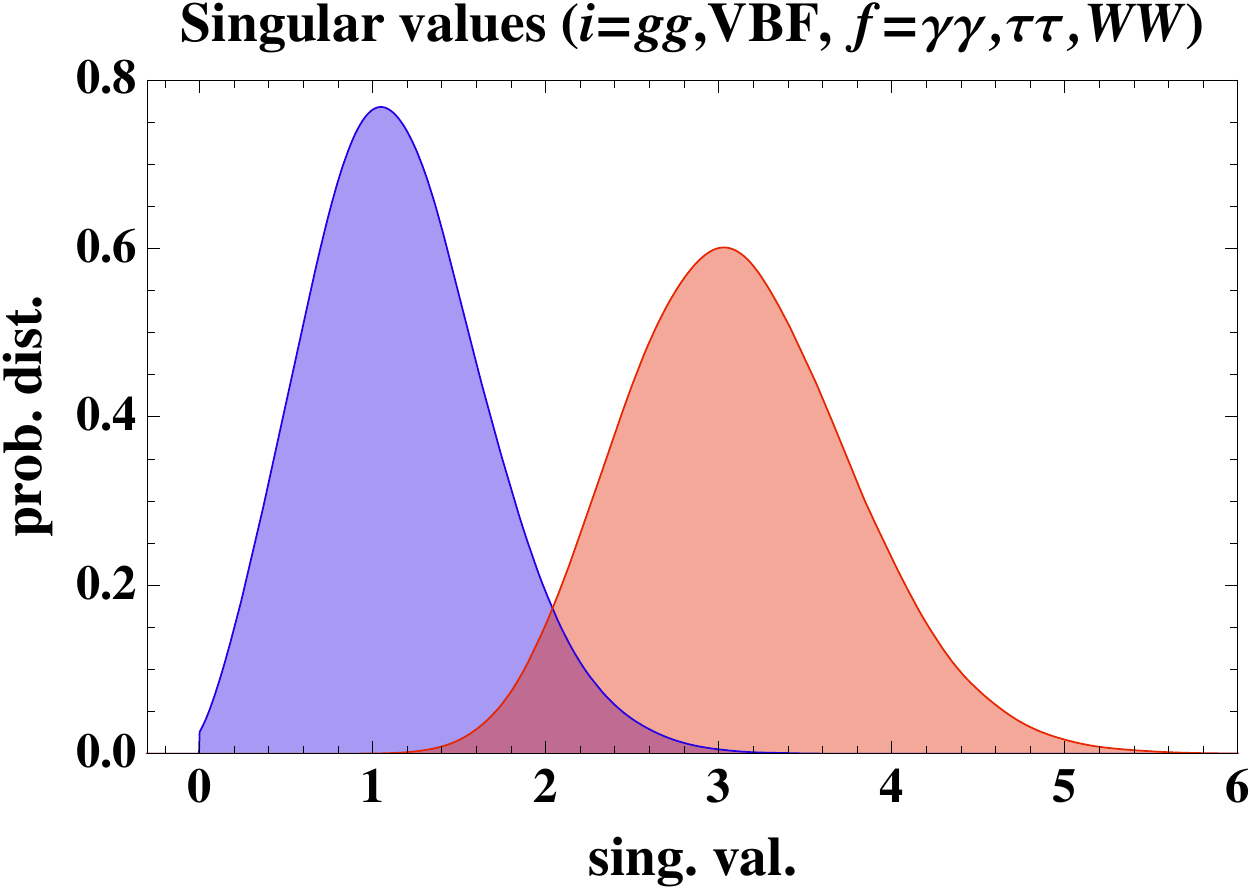}
\caption{\small {\it Left:} The probability distributions for the $(\Lambda_1)_{11}$ (blue) and $(\Lambda_1)_{12}$ (red) entries with present data and for $10\times$ more data assuming present central values and the dominance of statistical errors. The $R_{if}$ matrix has entries from $i=\{gg,$VBF\} and $f=\{\gamma\gamma,\tau\tau,WW\}$ channels. {\it Right:} The probability distributions for the two singular values of the $2\times 3$ $R_{if}$ matrix.}
\label{fig:mat.el.val.distrib:2x3}
\end{center}
\end{figure}

%
Obtaining the rank of $R$ may be done most generally by using
singular value decomposition (SVD),
\beq
   R = U S V^T,
   \label{eq:SVD}
\eeq
where $U$ and $V$ are orthogonal matrices with dimensions $n_i$ and $n_f$
respectively, and $S$ is a rectangular ($n_i\times n_f$) diagonal matrix.
The diagonal elements of $S$ are the singular values,
$\lambda_i\equiv S_{ii}$ (no summation), and may be obtained by
diagonalizing $RR^T$ (using $U$) or $R^TR$ (using $V$).
By definition $\lambda_i$ are non-negative, which can always be accomplished by appropriate redefinition of $U$ and $V$.  It is convenient to define $S$, $U$
and $V$, such that the singular values are ordered with $\lambda_1$ the largest.
The rank of an exactly known matrix is given by the number of nonzero singular
values $N(\lambda_i\neq 0)$. Since in our case the matrix $R$ is only known to a certain precision,
these singular values have distributions induced by the uncertainties
of the matrix elements.
To obtain a statistical test of $\rkR$ we follow the procedure of~\cite{Kleibergen:2006}.
We split $U$, $S$ and $V$ into the block matrices
\beq
U=\begin{pmatrix}
U_{r\times r} & \dots\\
\dots  & \tilde U
\end{pmatrix}, 
\qquad
S=\begin{pmatrix}
S_{r\times r} & 0\\
0 & \tilde S
\end{pmatrix},
\qquad
V=\begin{pmatrix}
V_{r\times r} & \dots\\
\dots  & \tilde V
\end{pmatrix},
\label{rxrdecomp}
\eeq
where $\tilde U$ is an $(n_i-r)\times(n_i-r)$ square matrix,
$\tilde S$ an $(n_i-r)\times(n_f-r)$ rectangular diagonal matrix, and
$\tilde V$ and $(n_f-r)\times (n_f-r)$ square matrix. We then construct 
\beq
\Lambda_r\equiv \tilde U \tilde S \tilde V^T,
\eeq
which is an $(n_i-r)\times(n_f-r)$ matrix that is zero if the singular
values are zero. (Note that we differ in the normalization of
$\Lambda_r$ from \cite{Kleibergen:2006}.) We can then test whether
singular values are consistent with zero simply by asking whether
$\Lambda_r$ is consistent with zero. We neglect
correlations and assume that the errors
for the matrix elements of $\Lambda_r$ are Gaussian, which we denote
by $\sigma[(\Lambda_r)_{ij}]$. We then define 
\beq
\chi^2(\Lambda_r)\equiv \sum_{i,j}\left(\frac{(\Lambda_r)_{ij}}{\sigma[(\Lambda_r)_{ij}]}\right)^2,
\label{SVD:chi2}
\eeq
which approaches the 
$\chi^2$ distribution for $(n_i-r)\times(n_f-r)$ degrees of freedom in
the large statistics limit.
While we do not propagate the correlations, one could
easily modify the above definition of $\chi^2(\Lambda_r)$ to take into 
account the correlations.
On average we expect $\chi^2(\Lambda_r)=(n_i-r)\times(n_f-r)$. Much
higher values would correspond to $\Lambda_r$ significantly nonzero,
with CL obtained from $\chi^2$ distribution, signaling that the rank
of the matrix $R$ is at least $r+1$. 
The advantage of working with $\Lambda_r$ instead of with singular values is that the entries of $\Lambda_r$ are not bound to be non-negative (cf. Fig. \ref{fig:mat.el.val.distrib:2x3}) so that it is easier to devise such a simple $\chi^2$ test.

%


Next, we apply the above method to the data. 
The largest matrix $R_{if}$ that we can form using present
measurements is $3\times 3$, with $i=\{gg,$VBF,VH\} and
$f=\{\gamma\gamma,\tau\tau,WW\}$ collected in Table
\ref{tab:rates}. 
Since the last row is least well measured we choose to use the $2\times3$ rectangular
submatrix  $i=\{gg,$VBF\} and
$f=\{\gamma\gamma,\tau\tau,WW\}$ for the numerical examples (we will return to the $3\times3$ case below).
The most obvious test is for $\rkR\geq1$, choosing
$r=0$ in \eqref{rxrdecomp}. This means that $\Lambda_0=R$. Neglecting
correlations we have $\chi^2(\Lambda_0)=36.1$ for $6$ d.o.f., which
corresponds to a probability $1-{\rm CL}=3 \cdot 10^{-6}$ that this is
just a statistical fluctuation. The signal significance is $4.7\sigma$
and we confirm the hypothesis of at least one resonance. 

Testing whether $\rkR=2$ we take $r=1$, so that $\Lambda_1$ is a
$1\times 2$ matrix. The probability distributions for the two entries
in $\Lambda_1$ are shown in Fig.~\ref{fig:mat.el.val.distrib:2x3} (left) and
were obtained from simple Monte Carlo with $10^6$ iterations assuming
Gaussian errors for the $R_{if}$ matrix elements. This gives
$\chi^2(\Lambda_1)=0.8$ which is not statistically significant for 2
d.o.f. as it corresponds to 
$0.4\sigma$ signal
significance. That is, with the current data we cannot test for more
than one resonance. With ten times more data, however, assuming the
same central values and scaling of errors as $1/\sqrt{L}$, one would
obtain a signal for  $\rkR=2$ with $3.9\sigma$ significance. While
this is just a very crude estimate, we conclude that with roughly an
order of magnitude more data we will be able to probe for a rank two matrix.

\subsection{Using The Determinant}

While the method described above is very general, in some cases we can 
devise simpler tests. When $R$ is  
a square block matrix ($n_i=n_f=n$) one can check 
if $\detR$ is significantly nonzero, which would imply that the rank of $R$ is $n$.
For $2\times2$ blocks (\ie, $n_i=n_f=2$), the determinant already
contains the maximal information available on the number of resonances, since
$\detR=0$ implies that one resonance suffices, while
a $\detR\neq0$ implies $n_r\geq2$.
For larger square blocks, however, the determinant can only distinguish
the $n_r\geq n$ case against $n_r< n$.
We will now give examples for both $2\times 2$ and $3\times3$ blocks.
While the former probes two resonances, the latter allows to probe three
resonances and interference.

First, consider the two $2\times 2$ sub-matrices
$R^A=(\{gg,\mbox{VBF}\},\{\gamma\gamma,\tau\tau\})$ and
$R^B=(\{gg,\mbox{VBF}\},\{\gamma\gamma,WW\})$.
The SM is excluded if any $R_{i,f}\ne 1$. The single resonance
hypothesis is excluded if  $\detR^A\ne 0$ or $\detR^B\ne 0$.
When formulated in term of the ratios
\beq 
r_1 \equiv
\frac{R_{{\rm VBF},WW}}{R_{{\rm VBF},\gamma\gamma}},\qquad r_2 \equiv
\frac{R_{gg,WW}}{R_{gg,\gamma\gamma}},\qquad r_3 \equiv \frac{R_{{\rm
      VBF},\tau\tau}}{R_{{\rm VBF},\gamma\gamma}}, \qquad r_4 \equiv
\frac{R_{gg,\tau\tau}}{R_{gg,\gamma\gamma}},
\eeq
these inequalities become $r_1\ne r_2$ or $r_3\ne r_4$. 
Ratios such as above have been considered
in~\cite{Gunion:2012he}, and have the virtue that
certain systematic errors cancel (this has been recently pointed out also in~\cite{Low:2012rj}).
As an example, consider a situation in which all the rates remain
at their current central values with much reduced errors,
except $R_{{\rm VBF}\,,WW}$,
which we take to be small instead of the current negative value:
\beq
   R_{gg\,,\gamma\gamma} = 1.6,\qquad
   R_{{\rm VBF}\,,\gamma\gamma} = 2.1,\qquad
   R_{gg\,,WW} = 0.8,\qquad
   R_{{\rm VBF}\,,WW} = 0.1\label{eq:rates1}.
\eeq
Interpreting such hypothetical results as coming from a single on-shell particle
is inconsistent, since the resulting signal matrix has nonzero determinant,
$\detR \neq0$.
In terms of ratios, this fact is evident in the inequality of the ratios
\beq
   \frac{R_{{\rm VBF}\,,WW}}{R_{{\rm VBF}\,,\gamma\gamma}}
  = 0.048,\qquad  \frac{R_{gg\,,WW}}{R_{gg\,,\gamma\gamma}}=0.5,
   \label{eq:ratios}
\eeq
while for a single resonance the two ratios would be identical and equal
to $\hat Br_{WW}/\hat Br_{\gamma\gamma}$.
%
If this inequality becomes certain in the future, it will therefore point to
$\rkR=2$, implying the existence of at least two mass
degenerate resonances, $\phi_1$ and $\phi_2$.
Such hypothetical result could be explained, for example if $\phi_1$ does
not couple to $W$ bosons while $\phi_2$ does not couple to photons.
%
%
Note that since we have
assumed low $R_{{\rm VBF},WW}$, if the resonances do not interfere, some of
the parameters need to be suppressed, for example, the ``Higgs'' coupling
to $WW$.  This is because in that case the $R_{if}$ values are sums of positive numbers. In contrast, no such suppression is needed for interfering resonances where the suppression can come from negative interference.
This provides more model building freedom.

A more realistic discussion includes experimental errors, where one tests for
a statistically significant deviation from $\detR=0$.
In Fig.~\ref{fig:contours:future}, we show contours of the integrated
luminosity required to exclude at $3\sigma$ the SM (blue) or the
single resonance hypothesis (green), in the $(r_1,r_2)$ and
$(r_3,r_4)$ planes. In the extrapolation we assume statistical scaling
of errors, that is $\propto 1/\sqrt{L}$, and central values taken from
\tabref{rates}.
We find that if the current central values remain where they are now,
then ${\mathcal O}(300 \ifb)$ at LHC 14TeV would suffice to exclude at
$3\sigma$ the single resonance hypothesis.

\begin{figure}[t]
\begin{center}
\includegraphics[scale=0.55]{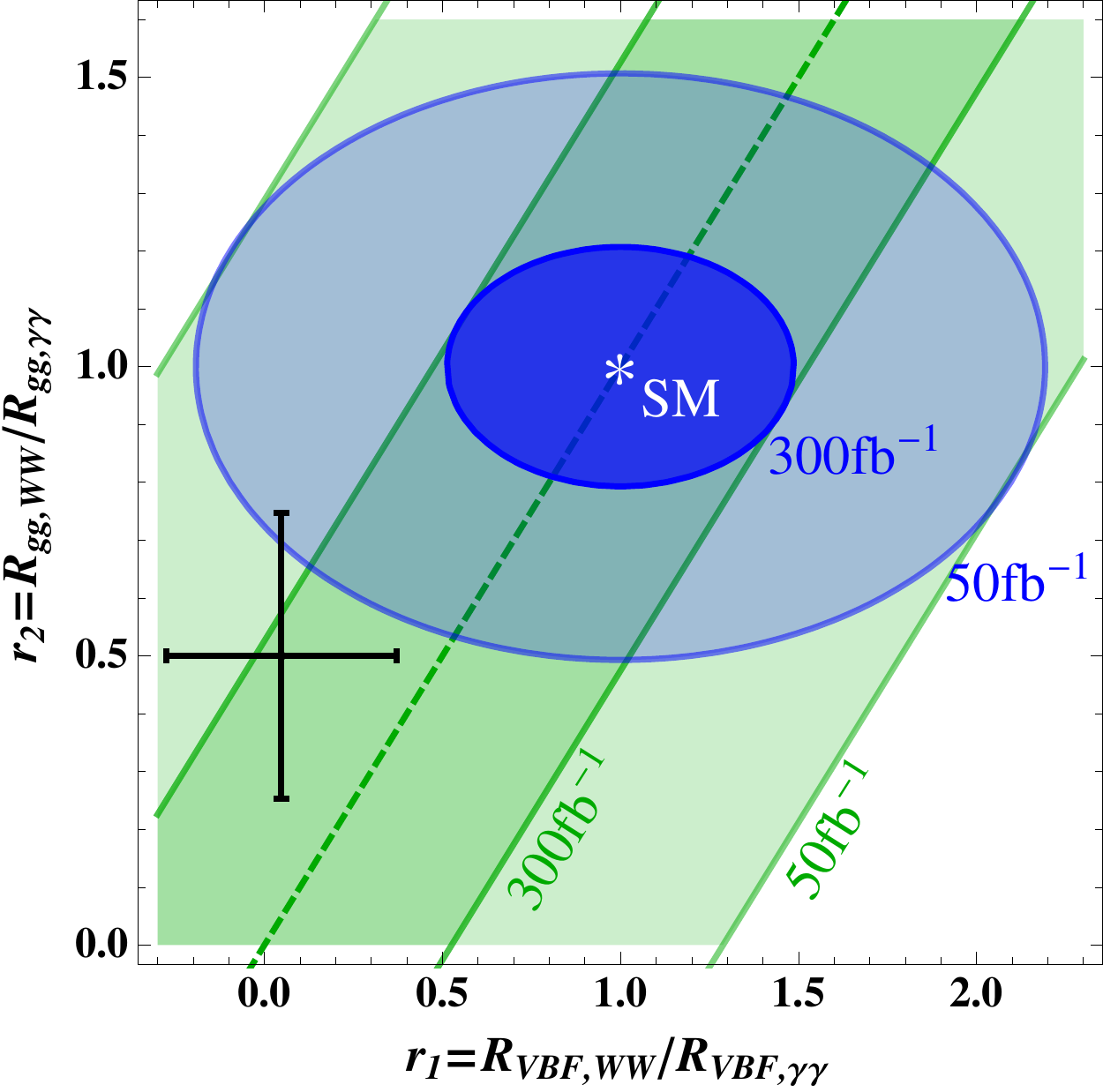}
~~~~~
\includegraphics[scale=0.57]{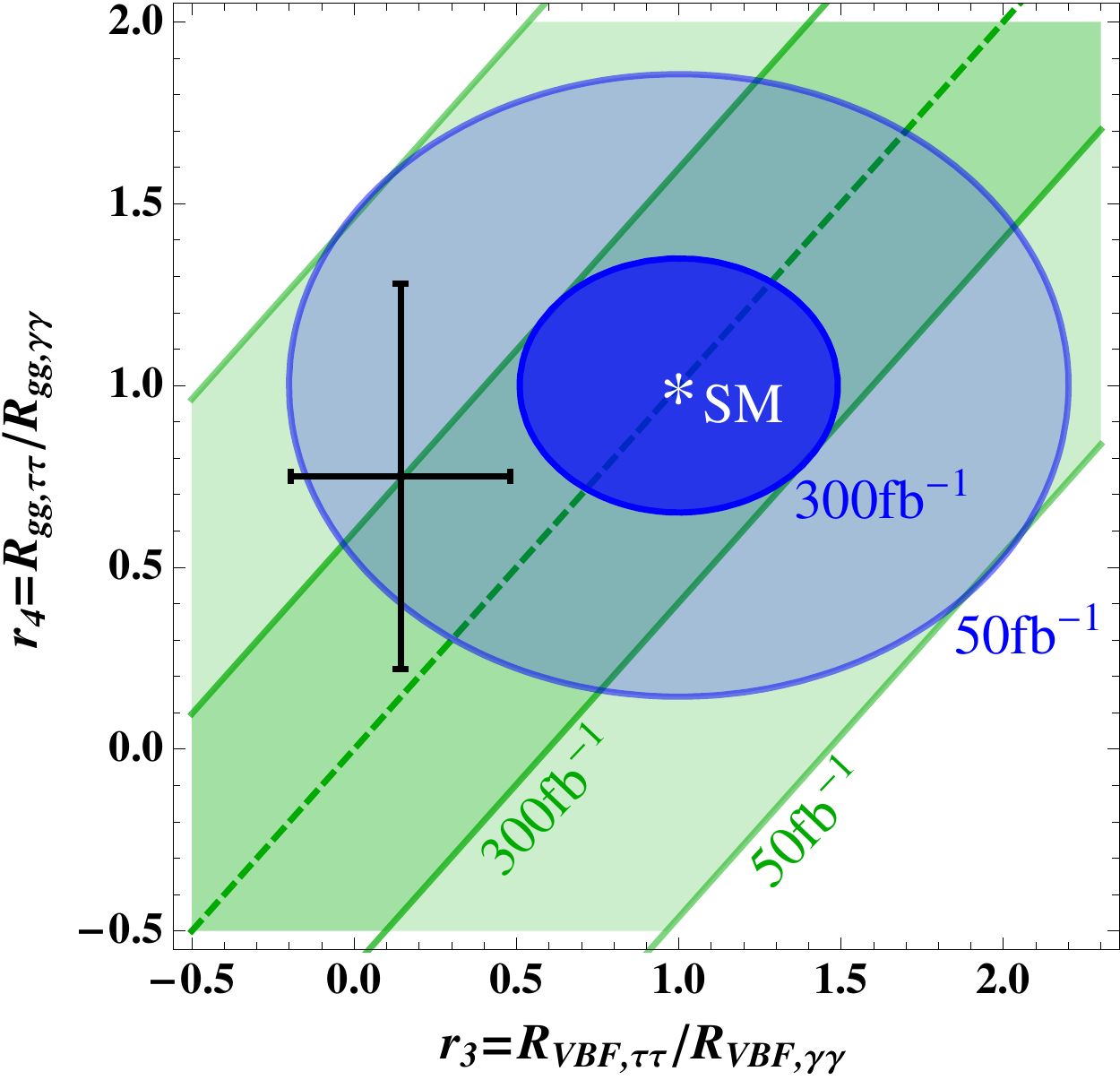}

\caption{\small Contours of constant integrated luminosity at LHC 14 TeV required to exclude
at $3\sigma$ the SM (the single resonance hypothesis) if the measurement is within blue (green) regions
in the $(r_1,r_2)$ plane (left) and in the $(r_3,r_4)$ plane (right). We have
assumed statistical scaling of errors, while
present $1\sigma$ measurements are denoted with black error bars.
The 14 TeV run projections were obtained by rescaling the
current $7+8$ TeV errors  with the increase in the higgs production cross section  \cite{xsecs}
 to reflect the gain in statistics.
%
\label{fig:contours:future}
}
\end{center}
\end{figure}



Finally, we discuss the possibility of testing for rank 3.
Note that this is not only a generalization from 2 resonances
to 3, it is also the lowest possible rank which can probe interference.
As an example we test for rank 3 using $\detR\ne0$ test for $n=3$. 
This  distinguishes between $\rkR=3$ and $\rkR=1,2$ cases.
The $\rkR=1$ and $\rkR=2$ cases may be resolved using
SVD, as discussed in Eqs. \eqref{eq:SVD}-\eqref{SVD:chi2}.
%
For inputs we use the 
$(\{gg,\mbox{VBF},VH\},\{\gamma\gamma,WW,\tau\tau\})$ block in Table \ref{tab:rates}.

First, consider an idealized future scenario, where all the central
values are as in~\tabref{rates}, except the negative ones, which we
take to be $+0.1$, for illustration purposes.  The relevant matrix
has rank 3, and can be decomposed as usual using three
non-overlapping resonances.  The same matrix can also result from 2 resonances,
provided they interfere (cf.~\secref{infinite}).  For example, the two resonances may have the
following amplitudes, 
\beqa 
\vec{A}_{gg} &=&
\lt(1,0.458\,e^{0.5i}\rt),\, \vec{A}_{\rm VBF} =
\lt(0.673\,e^{-1.5i},1.088\,e^{0.4i}\rt),\,
\vec{A}_{VH} = \lt(1.11\,e^{0.05i},1.277\,e^{2i}\rt),\nonumber\\
\vec{A}_{\gamma\gamma} &=& \lt(1,1.95\,e^{1.2i}\rt),\, \vec{A}_{WW^*}
= \lt(0.9\,e^{1.9i},0.69\,e^{-3.1i}\rt),\, \vec{A}_{\tau\bar\tau} =
\lt(1.1\,e^{0.4i},0.17\,e^{1.6i}\rt).  
\eeqa 
Note that this example requires a non-universal phase.  As we mentioned before,
an example with a universal phase does not always exist.  However, often a
universal phase model can produce a matrix that is in the vicinity of the
one that is needed. Thus, it is not clear how much data one would need
in order to probe non-universal phases.


We are now ready to discuss the more realistic case with finite statistics.
We show the projected sensitivity in Fig. \ref{fig:3states}({\it left}), 
assuming statistical scaling of errors, and assuming the cross sections scale
with center of mass energy as they do in the SM.
In this case ${\mathcal O}(300 \ifb)$ at LHC 14 TeV are needed for $3\sigma$ 
evidence of two resonance case using the $\rkR=2$ test described above. We estimate that   ${\mathcal O}(2000 \ifb)$ at LHC 14 TeV are needed for $3\sigma$ 
evidence of three overlapping resonances using the test that $\det R\ne 0$.
It is important to note that all these estimates crucially depend on the input parameters and ${\mathcal O}(1)$ change in the central value of just one parameter can lead to a factor of few smaller required luminosities.
Therefore, in the right panel we show a plot where the central values are allowed
to vary.
Contours of constant luminosity are shown where the significance for
discovering 3 non-interfering or 2 interfering resonances ($\detR\neq0$)
reaches $3\sigma$. The error has been propagated linearly by expanding the determinant.
While this is clearly not adequate for large relative errors, it provides a
rough estimate for the significance.

\begin{figure}[t]
\begin{center}
\includegraphics[scale=0.67]{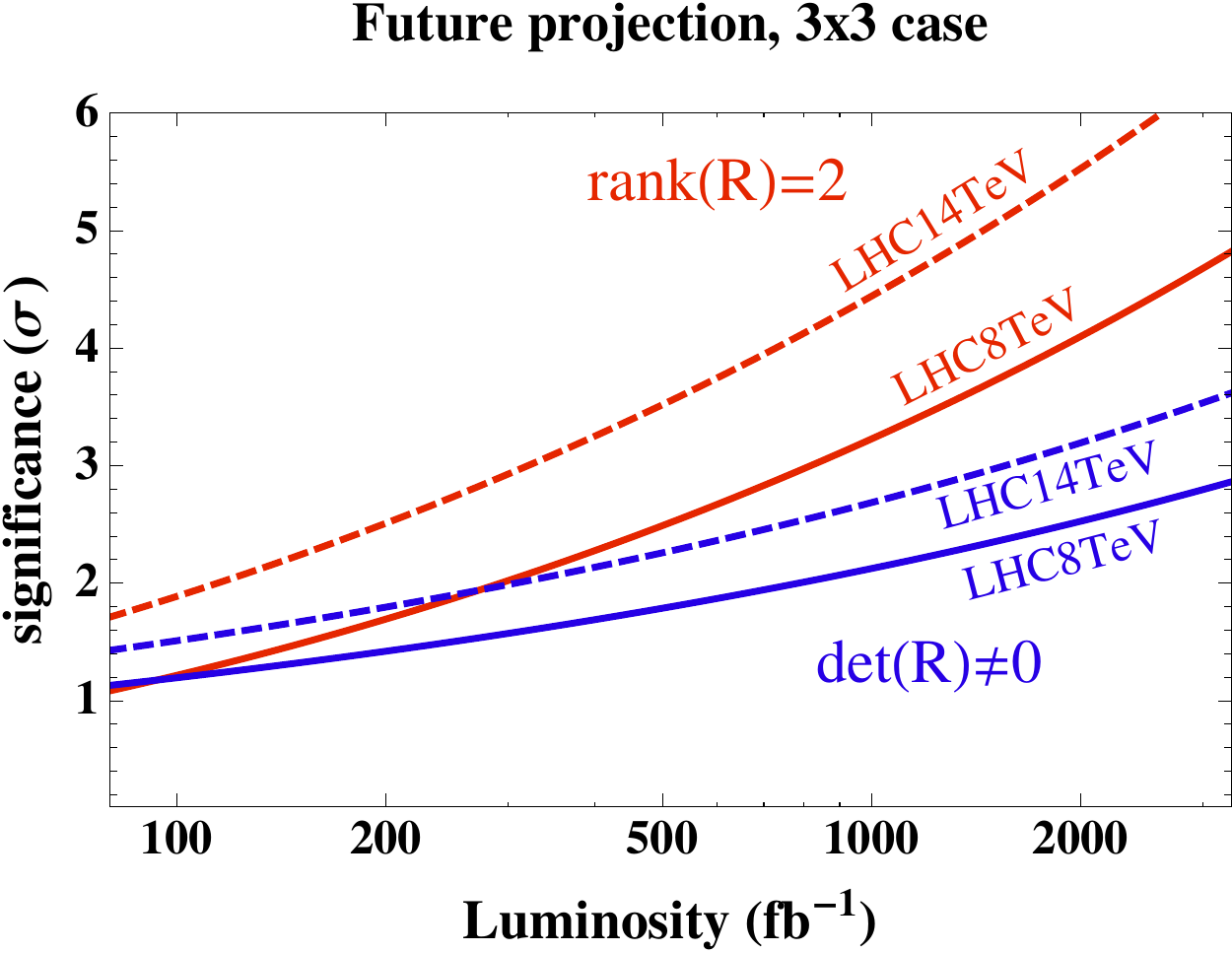}
~~~~~
\includegraphics[scale=0.54]{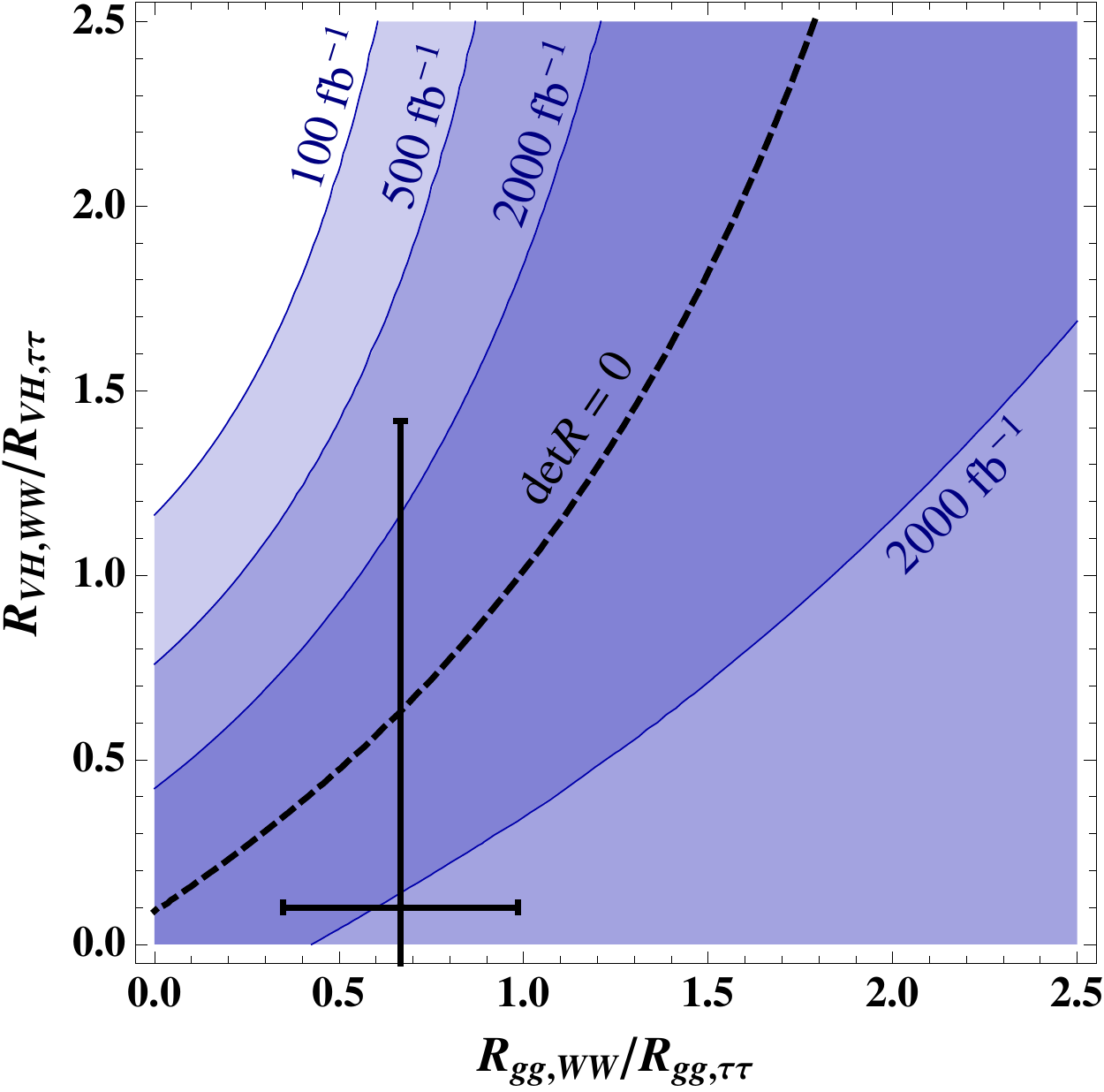}
\caption{\small 
{\it Left:} 
The projected significance for a discovery of 3 (2) overlapping resonances using $3\times 3$ $R$ matrix (see text) using the $\detR\ne0$ ($\rkR\geq 2$) tests are shown with solid and dashed blue (red) lines for LHC at 8TeV and LHC at 14 TeV, respectively, as a function of integrated luminosity.
{\it Right:}
Integrated luminosity at 14 TeV LHC required to probe 3 non-interfering
or 2 interfering resonances with $3\sigma$ significance as a function of the ratios
$R_{gg,WW^*}/R_{gg,\tau\bar\tau}$ and $R_{VH,WW^*}/R_{VH,\tau\bar\tau}$,
keeping $R_{gg,WW^*}$ and $R_{VH,\tau\bar\tau}$ fixed to their central
values in~\tabref{rates}.
\label{fig:3states}}
\end{center}
\end{figure}

\section{Summary and Outlook}
We have shown that one can test for the presence of multiple resonances
in the recently observed Higgs signal, even if the multiple resonances
are degenerate in mass with respect to the experimental
resolution. The test requires measuring algebraic properties of
the event rate matrix, $R$, in particular its rank.

For purposes of the current Higgs signal at the LHC, we find that if the current
central values remain 
then ${\mathcal O}(100\ifb)-{\mathcal O}(300 \ifb)$ of integrated luminosity at the 14 TeV LHC  may be sufficient to detect evidence for 2 resonances with $3\sigma$ significance.
Detecting  three resonances (or interference between two) would
require more data,  ${\mathcal O}(2000 \ifb)$. Of course, the
situation can dramatically change based on how
the central values move. 

While in this work we pointed out that the rank of the $R$ is an
important observable, we did not carry the full statistical treatment
of the available data. This task must be performed by the experimental
collaborations, taking into account correlations and systematics, and correctly
accounting for the propagation of error.

The principles we have discussed in this work are universal and may prove
suitable not just for Higgs physics at the LHC.  For example, new resonance peaks
may be found in the following years, as motivated by many extensions of the SM,
and investigating their signal matrices might be the only way to probe degeneracy,
especially in a hadron collider where the mass resolution cannot be very good.
%



\acknowledgments
We thank Haim Bar, Kfir Blum, Israel Klich and Carlos Wagner for valuable discussions, and the
Aspen Center for Physics (supported by the NSF grant PHY-1066293),
where this work has been initiated.
ZS is supported in part by the National Science Foundation under Grant
PHY-0969739. JZ acknowledges the hospitality of LBNL High Energy Theory Group where part of this project was completed.


\begin{thebibliography}{99}
\bibitem{:2012gk} 
  G.~Aad {\it et al.}  [ATLAS Collaboration],
  Phys.\ Lett.\ B {\bf 716}, 1 (2012)
  [arXiv:1207.7214 [hep-ex]].
  
\bibitem{:2012gu} 
  S.~Chatrchyan {\it et al.}  [CMS Collaboration],
  Phys.\ Lett.\ B {\bf 716}, 30 (2012)
  [arXiv:1207.7235 [hep-ex]].

\bibitem{Batell:2012mj} 
  B.~Batell, D.~McKeen and M.~Pospelov,
 arXiv:1207.6252 [hep-ph].  

\bibitem{Gunion:2012he}
  J.~F.~Gunion, Y.~Jiang and S.~Kraml,
 arXiv:1208.1817 [hep-ph].  
 
\bibitem{Gunion:2012gc}
J.~F.~Gunion, Y.~Jiang and S.~Kraml,
Phys.\ Rev.\ D {\bf 86} (2012) 071702
[arXiv:1207.1545].

\bibitem{Ferreira:2012nv} 
  P.~M.~Ferreira, H.~E.~Haber, R.~Santos and J.~P.~Silva,
  arXiv:1211.3131 [hep-ph].
  
\bibitem{Drozd:2012vf}
A.~Drozd, B.~Grzadkowski, J.~F.~Gunion and Y.~Jiang,
arXiv:1211.3580 [hep-ph].

\bibitem{Ellis:2005fp}
J.~R.~Ellis, J.~S.~Lee and A.~Pilaftsis,
Phys.\ Rev.\ D {\bf 71} (2005) 075007
[hep-ph/0502251].

\bibitem{Ellis:2004fs}
J.~R.~Ellis, J.~S.~Lee and A.~Pilaftsis,
Phys.\ Rev.\ D {\bf 70} (2004) 075010
[hep-ph/0404167].


\bibitem{Dixon:2003yb} 
  L.~J.~Dixon and M.~S.~Siu,
  Phys.\ Rev.\ Lett.\  {\bf 90}, 252001 (2003)
  [hep-ph/0302233].

\bibitem{Martin:2012xc} 
  S.~P.~Martin,
  Phys.\ Rev.\ D {\bf 86}, 073016 (2012)  [arXiv:1208.1533 [hep-ph]].  

\bibitem{Low:2012rj} 
  I.~Low, J.~Lykken and G.~Shaughnessy,
  Phys.\ Rev.\ D {\bf 86}, 093012 (2012)
  [arXiv:1207.1093 [hep-ph]].
  
\bibitem{Bertolini:2012gu} 
  D.~Bertolini and M.~McCullough,
  arXiv:1207.4209 [hep-ph].  

\bibitem{Freitas:2012kw}
A.~Freitas and P.~Schwaller,
arXiv:1211.1980 [hep-ph].

\bibitem{ATLAStautau}
The ATLAS Collaboration, ATLAS-CONF-2012-160.

\bibitem{CMStautau}
The CMS Collaboration, CMS-HIG-12-043. 

\bibitem{ATLASbbbar}
The ATLAS Collaboration, ATLAS-CONF-2012-161; 
The CMS Collaboration, CMS-HIG-12-044. 

\bibitem{CMSttbar}
The CMS Collaboration, CMS-HIG-12-025. 

\bibitem{ATLASgamma}
The ATLAS Collaboration, ATLAS-CONF-2012-168; 
The CMS Collaboration, CMS-HIG-12-015. 

\bibitem{ATLAScomb}
The ATLAS Collaboration, ATLAS-CONF-2012-127. 

\bibitem{CMScomb}
The CMS Collaboration, CMS-HIG-12-045. 

\bibitem{ATLASZZ}
The ATLAS Collaboration, ATLAS-CONF-2012-092; 
The CMS Collaboration, CMS-HIG-12-016. 

\bibitem{Espinosa:2012ir}
J.~R.~Espinosa, C.~Grojean, M.~Muhlleitner and M.~Trott,
JHEP {\bf 1205} (2012) 097
[arXiv:1202.3697 [hep-ph]];
J.~R.~Espinosa, C.~Grojean, M.~Muhlleitner and M.~Trott,
arXiv:1207.1717 [hep-ph].

\bibitem{Azatov:2012bz}
A.~Azatov, R.~Contino and J.~Galloway,
JHEP {\bf 1204} (2012) 127
[arXiv:1202.3415 [hep-ph]].

\bibitem{Azatov:2012qz}
A.~Azatov and J.~Galloway,
arXiv:1212.1380 [hep-ph].

\bibitem{Carmi:2012yp}
D.~Carmi, A.~Falkowski, E.~Kuflik and T.~Volansky,
JHEP {\bf 1207} (2012) 136
[arXiv:1202.3144 [hep-ph]];
D.~Carmi, A.~Falkowski, E.~Kuflik, T.~Volansky and J.~Zupan,
JHEP {\bf 1210} (2012) 196
[arXiv:1207.1718 [hep-ph]].

\bibitem{Kleibergen:2006}
Kleibergen, F. and  Paap, R., Journal of Econometrics,
Vol. 133, Issue 1, July 2006, 97

\bibitem{xsecs}
  LHC Higgs Cross Section Working Group, S.~Dittmaier, C.~Mariotti, G.~Passarino, and R.~Tanaka (Eds.), 
  {\sl Handbook of LHC Higgs Cross Sections: 1. Inclusive Observables}, 
  CERN-2011-002 (CERN, Geneva, 2011), {\tt arXiv:1101.0593 [hep-ph]}.


\end{thebibliography}
\end{document}